\documentclass[vecphys]{svmult}

\usepackage{makeidx}         
\usepackage{graphicx}        
\usepackage{multicol}        
\usepackage[bottom]{footmisc}

\makeindex             

\begin{document}

\title*{Observational Cosmology with the ELT and JWST}
\author{Massimo Stiavelli\inst{1}}
\institute{Space Telescope Science Institute, 3700 San Martin Dr., Baltimore MD21218, USA
\texttt{mstiavel@stsci.edu}}
%
%
\maketitle


\section{Dark areas in observational cosmology}
\label{sec1}

The three hottest themes in modern observational cosmology are all qualified by the word "dark": Dark Matter, Dark Energy and Dark Ages. Dark Matter and Dark Energy are studied by a variety of techniques both from the ground and from space and we expect that the James Webb Space Telescope (JWST, \cite{GardnerSSR}) will contribute to these areas( e.g. \cite{RiessLivio2006}) but will not be dominant. In contrast, the study of the Cosmic Dark Ages is one of the four main themes of JWST and I expect its contributions to this field to be major.

The Cosmic Dark Ages \cite{LoebBarkana} are the epoch of cosmic history bracketed by the recombination of Hydrogen at redshift $z\sim1300$ and its reionization at $z\sim6-7$. The three most important milestones during the Dark Ages are the formation of the first (Population III) stars \cite{BrommLarson,TrentiStiavelliFS}, the formation of the first galaxies\cite{Abel}, and the reionization of Hydrogen\cite{Fan}. In the following section I will review our present observational knowledge in these areas.

\section{Observing the Dark Ages with present instrumentation}
\label{sec2}

The increasingly high optical depth shortwards of Lyman$\alpha$ in the spectra of SDSS QSOs at redshift $z\geq6$ \cite{Fan} is a strong indication that reionization is completed at $z\approx6$. This is also compatible with the 3 year WMAP Compton depth measurement \cite{Spergel,ShullVenkatesan}.
If reionization is completed at $z\approx6$ and if it is a fast process occurring over a $\Delta z\simeq 1$ we should be able to identify the galaxies responsible for it \cite{Stiavellietal04a}. Indeed, this was one of the primary motivations for the Hubble Ultra Deep Field (UDF, \cite{Beckwith}). 
Focusing on galaxies found with the Lyman Break technique (LBT), the Great Observatory Origins Deep Survey (GOODS, \cite{Giavalisco}) and the UDF 
have provided us with sample of $> 500$ i-dropout galaxies likely to be at $z\sim 6$ \cite{Bouwens06,Bouwens07}. Estimating their ionizing flux  involves an extrapolation from the observed, non-ionizing, UV continuum to the Lyman continuum. Similarly, uncertainties on the ionizing photons escape fraction and on the gas clumping (driving recombinations) render uncertain the required amount of radiation needed for reionization. The observed flux is tantalizing close to what is needed but the uncertainties have led different groups to diffent conclusions depending on whether one assumes the presence of large numbers of faint dwarf galaxies \cite{YanWindhorst}, or a top heavy stellar mass function and low metallicity \cite{Stiavellietal04b}, or galaxy parameters similar to those found at lower redshift \cite{Bunker}.

\begin{figure}
\centering
\includegraphics[height=5.5cm]{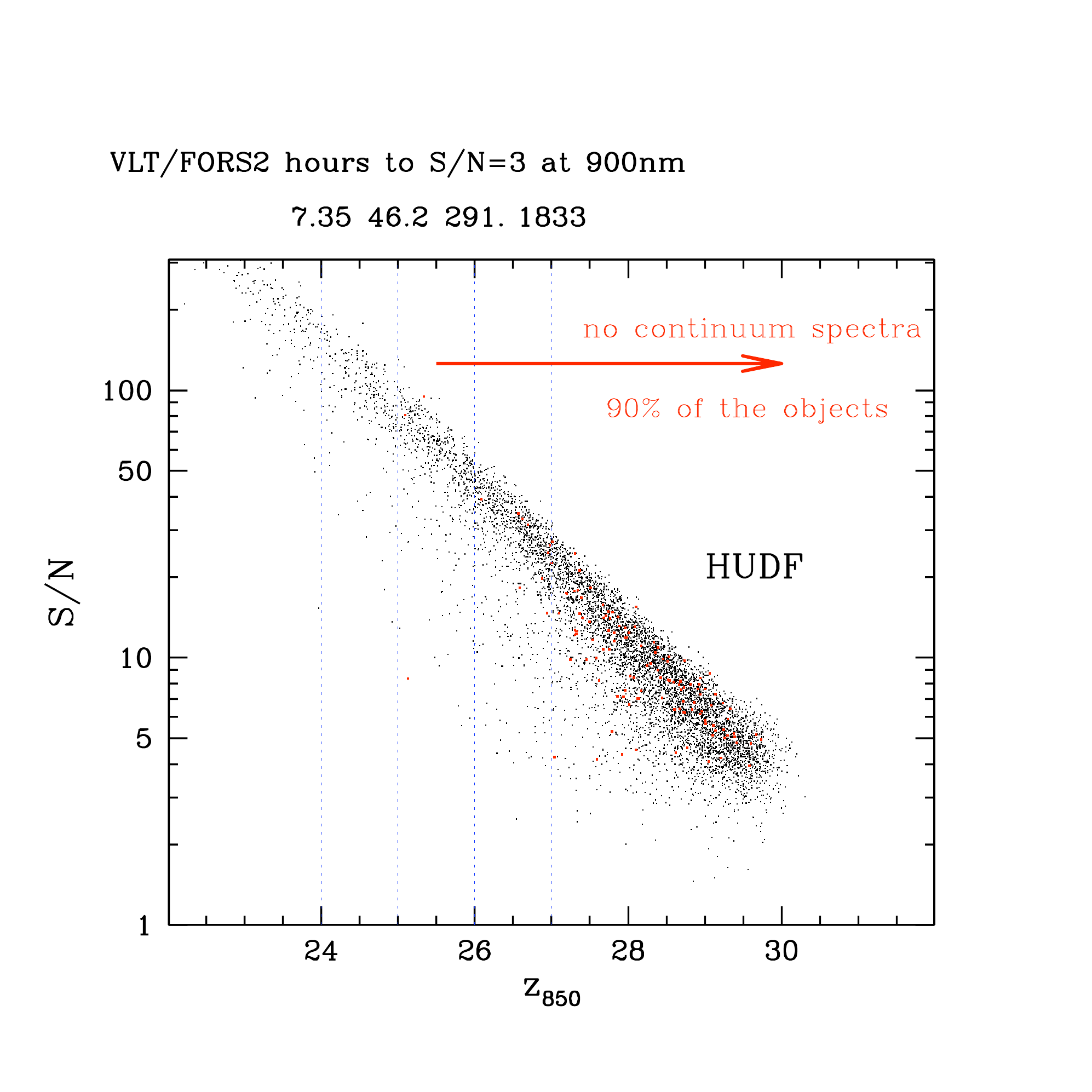} 
\includegraphics[height=5.5cm]{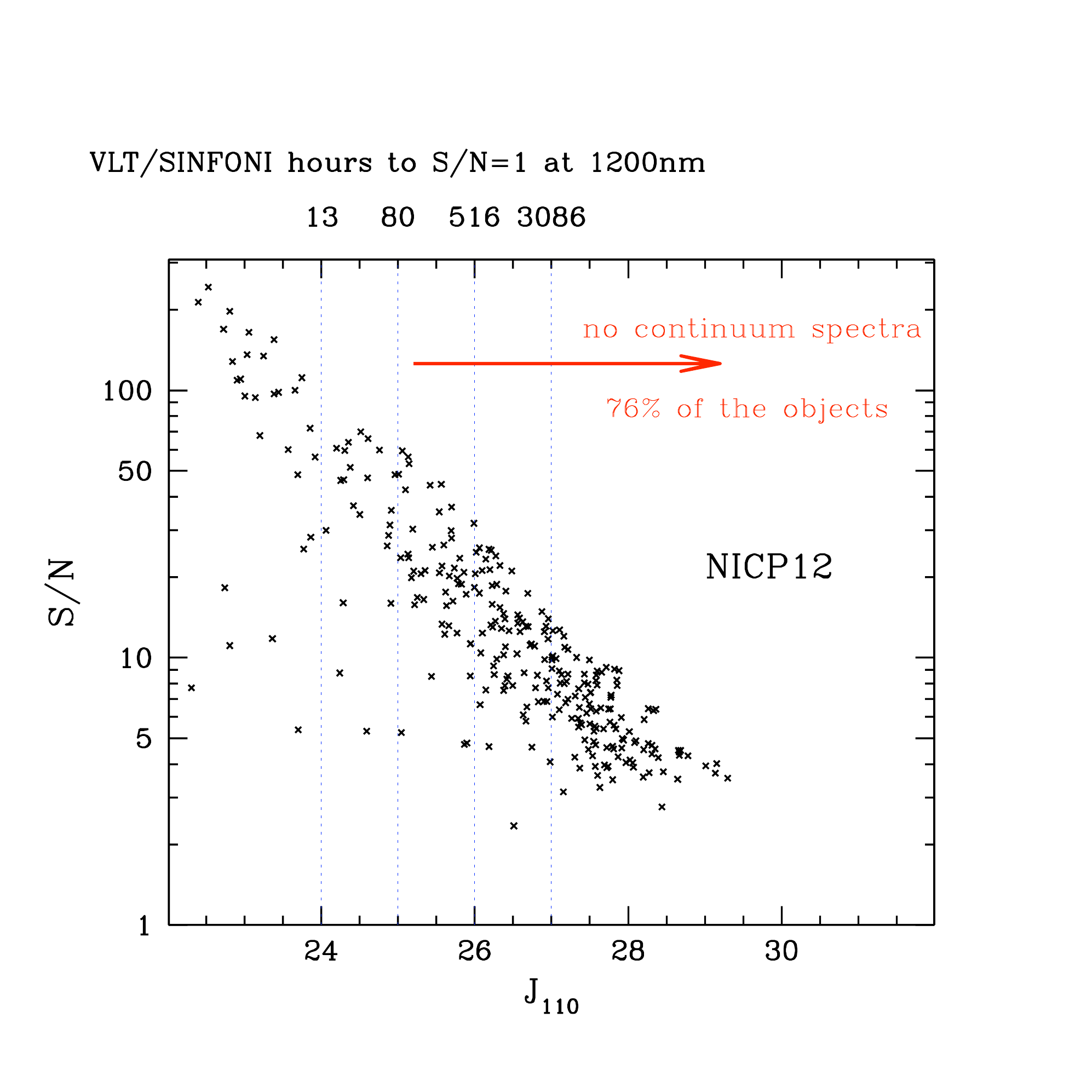}
%
%
\caption{Left: S/N vs $z_{850}$ magnitude for the UDF. Points in red represent i-dropout galaxies.  Right: S/N vs magnitude for the NICP12 field of the UDF followup program. For both we give the integration times for spectroscopic followup (see text).}
\label{fig1}       
\end{figure}

Whether or not the galaxies at $z\sim 6$ are sufficient to reionize the Universe has implications on the evolution of the luminosity function (LF) of galaxies. Indeed, if the LF evolves very fast at $z> 6$, it is unlikely that galaxies at $z>6$ will contribute much to reionization and those at $z\sim6$ have to do it by some combination of a steep, dwarf-rich LF and low metallicity. On the other hand, if there is rapid evolution of the LF at $z>6$, galaxies at $z>6$ will contribute to the ionizing background extending the duration of reionization and pushing the time of the formation of the first galaxies to higher redshift.

In order to search for galaxies at $z\geq7$ we started a UDF followup study \cite{Oeschetal}. The preliminary result is that there is a deficit of high-z objects and this has been first reported by \cite{BouwensrapidLF}. However, it is unclear whether the evidence of a rapid evolution of the LF is statistically significant once the systematic errors of the modeling and the impact of cosmic variance are properly included \cite{TrentiStiavelliCV}. Thus, the jury is still out on whether the LF is really evolving rapidly. It is well possible that surveys carried out with the Wide Field Camera 3 installed in the Hubble Space Telescope during the upcoming servicing mission will clarify this issue. Ongoing surveys exploiting gravitational lensing might also provide us with very high redshift objects \cite{StarkEllis} but deriving a LF  from them is going to be hard given the small effective volume and the related high cosmic variance. An alternative to the LBT  is the search for Lyman$\alpha$ emitting sources by use of the narrow-band excess technique which has delivered the highest redshift spectroscopically confirmed galaxy known to date\cite{Iye}.


%
%

However, even with well identified candidates we are left with two open questions: how far can one trust dropout selections (or single line redshifts) in an essentially unprobed redshift region and how can one decrease the uncertainty in the ionizing photons output by improved physical modeling of these galaxies. Both questions require spectroscopy to be addressed.

\section{The need for spectroscopy}
\label{sec3}

The majority of the galaxies we are interested in at $z \geq 6$ are beyond the reach of spectroscopic study from the ground. This is illustrated in Figure \ref{fig1}.
In the left panel  we show the $z_{850}$ magnitude vs S/N plot for  the UDF. Points in red are the i-dropout galaxies. As an illustration of the difficulty to obtain high S/N spectra at this faint levels, I report in the Figure the integration time (in hours) with FORS2 on the Very Large Telescope (VLT) required to achieve a S/N=3 per resolution element at 900nm, which is a representative wavelength for the continuum below Lyman $\alpha$ at $z<6.4$. Here we focus on the continuum because about 50\% of Lyman break galaxies are expected not to have Lyman $\alpha$ in emission \cite{Steidel} and in any case studying the fainter lines needed for deriving the physical properties of these objects would require at least this level of S/N. The Figure shows that about 90\% of the galaxies in the UDF and the vast majority of i-dropout galaxies would require 100 hours of integration or more with the VLT. For a few objects, single emission line redshifts can be obtained at fainter magnitudes but between the faintest objects at S/N=5 and $z_{850} = 29.5$ and those for which one can typically obtain a useful spectrum we have a gap of 4.5 magnitudes and even for shallower (but wider) surveys like GOODS there is a gap of at least 2 magnitudes.

A similar situation is present in the near-IR and is illustrated in the right panel of Figure \ref{fig1} where I plot the S/N vs $J_{110}$ magnitude for the NICP12 field in the UDF area \cite{Oeschetal}. Here the spectroscopic benchmark is an integration with SINFONI on the VLT with S/N=1 in the continuum at 1200nm. I have adopted a lower S/N because when studying the continuum one could presumably rebin several resolution elements after having masked out the brightest OH lines.  Using 100 hours as the practical limit, the Figure shows that objects fainter than $J_{110} = 25.2$ are too faint for spectroscopic study and this implies a gap of about 3 magnitudes and includes 76 \% per cent of the galaxies in the NICP12 field. The installation of WFC3 on HST will further increase this gap.

JWST will be able to obtain spectra of many galaxies using the multi-object spectrometer NIRSpec. However, the field-of-view (FOV) of NIRSpec is only about 10 square arcmin. This is well matched to the size of the UDF,  but NIRSpec would require at least 30 separate integrations to cover GOODS. Morover, JWST will be able to obtain images with NIRCam deeper by a couple of magnitudes than the UDF thus preserving a gap between imaging and followup spectroscopy.

\section{Rarity of High-z galaxies}
\label{sec4}

The first galaxies, almost by definition, would be rare. Clearly we don't yet know how rare, how faint, and at what redshift they will occur. Using a simple model, we have derived expected counts for high-z dropouts objects seen by NIRCam \cite{TrentiStiavelliCV}. Based on that prediction and considering the FOV size of about 9.7 square arcmin for NIRCam one would expect about 10 objects per deep NIRCam field at $z\simeq10$, about  1 per field at $z\simeq12$ and about 0.1 per field at $z\simeq15$.

Moving from first galaxies to first stars also forces us to contemplate rare objects. Indeed it is likely that we will be able to detect Population III stars only when they produce ultra-bright pair-instability supernovae and such supernovae are probably extremely rare, e.g., 4 per square degree per year at $z=15$ \cite{WeinmanLilly}. In order to detect them, one would need degree-scale surveys with sensitivity around $AB=26-27$ in the near-IR. Clearly for these objects the limiting factor wouldn't be the JWST sensitivity but its field of view.

\section{How can the ELT help?}
\label{sec5}
JWST is optimized for IR imaging and low-medium resolution spectroscopy over a modest FOV. The need for identifying rare objects or emplying higher spectroscopic resolving power require ground based telescopes of the 30+m class like, e.g., the Extremely Large Telescope (ELT).

To help close the  gap between spectroscopy and imaging it would be very desireable to have a high-throughput multi-object spectrograph operating in the red and in the near-IR on the ELT with a field of view as large as practical and a resolving power between 3000 and 5000 so as to be able to reject the brightest OH lines and retain a low background between such lines. Alternatively one could work at lower resolution if efficient OH-suppression spectrographs become feasible. Full adaptive optics would not be very important for such an application as galaxies are marginally resolved at the resolution of HST and JWST and, unlike the case of stars, one would not gain a background reduction by having a sharper point spread function. To the contrary, in the detector limited case, perhaps achievable between the lines, one might lose S/N by over-resolving the galaxies. Thus, the two critical parameters for such a spectrograph would be throughput and multiplexing.

Kinematical study of faint high-z galaxies might well require higher resolving power (up to 10,000) than afforded by JWST due to their low mass \cite{StiavelliLiege} and this could be addressed by the ELT. Similarly, absorption line studies of QSOs at $z>7$ will need to be done from the ground at $R\sim 30,000$.

Helping JWST to identify Population III objects would require monitoring a large area (at least one square degree) down to $AB=26$ or 27. This may or may not be possible on the ground depending on the feasibility and efficiency of OH-suppression imagers and possibly moderate field MCAO imagers (as supernovae are point-like and one would benefit from a sharper point spread function). The belief that pair-instability supernovae evolve slowly (e.g. 200 days to decrease by 2 magnitudes from the peak),  combined with cosmological time dilation  makes a yearly monitoring program acceptable and would render imaging programs requiring many nights of integration feasible.

\index{paragraph}

%
%
%



\printindex
\end{document}